\documentclass[preprint,preprintnumbers,amsmath,amssymb]{revtex4}
\usepackage{graphicx}
\usepackage{dcolumn}
\usepackage{bm}

\begin{document}
\title{Switching speed distribution of spin-torque-induced magnetic
reversal}
\author{J. He$^a$, J. Z. Sun$^b$ and S. Zhang$^a$}
\affiliation{$a)$ Department of Physics and Astronomy, University of
Missouri-Columbia, Columbia, MO 65211 \\ $b)$ IBM T. J. Watson
Research Center, P. O. Box 218, Yorktown Heights, NY 10598.}

\begin{abstract}
The switching probability of a single-domain ferromagnet under
spin-current excitation is evaluated using the Fokker-Planck
equation(FPE). In the case of uniaxial anisotropy, the FPE reduces
to an ordinary differential equation in which the lowest eigenvalue
$\lambda_1$ determines the slowest switching events. We have
calculated $\lambda_1$ by using both analytical and numerical
methods. It is found that the previous model based on thermally
distributed initial magnetization states \cite{Sun1} can be
accurately justified in some useful limiting conditions.
\end{abstract}
\maketitle

\section{Introduction}

Fast and reliable nanosecond level writing is essential for
spin-torque-induced switching in memory and recording technologies.
One intrinsic source for write threshold distribution is thermal
fluctuation, two aspects of which could affect reversal time
$\tau_s$. First, the initial position of the magnetic moment is
thermally distributed at the time the reversal field or current is
applied, causing a variation in switching time. Secondly, during
reversal, thermal fluctuation would modify the orbit, causing
additional fluctuation for $\tau_s$ even for identical initial
conditions. A reversal time distribution due to a thermalized
initial condition was recently estimated \cite{Sun1}. Here we
present a Fokker-Planck formulation for the full problem including
both thermalized initial condition and reversal orbit with estimates
for the reversal time and its distribution.

Fokker-Planck equations have been successfully applied to macro-spin
systems with a static thermal distribution for the extraction of a
small thermal escape probability. For example, the distribution for a
thermally activated reversal has been analytically and numerically
obtained when the applied magnetic field is smaller than the coercive
field \cite{Brown,Aharoni,Scully,Coffey,Bertram}. There the switching
is through thermally assisted transition from one metastable energy
minimum to equilibrium. The ensemble-averaged $\tau_s$ depends
exponentially on the ratio between the reversal energy barrier and
thermal energy $k_B T$. These subthreshold results have recently been
extended to include the effect of spin-torque, which modifies the
effective temperature \cite{Zhang,Visscher,Sun2}.

For fast, dynamically driven switching, the applied spin-torque is
above the zero-temperature reversal threshold. For finite temperature,
this leads to a time-dependent Fokker-Planck equation in which the
time-evolution of the probability represents the thermalized reversal
of the macro-spin. Here we solve this problem in the colinear geometry
with a uniaxial anisotropy energy landscape and the presence of a spin
torque.

\section{Fokker-Planck equation}
Consider an ensemble time-dependent magnetization probability
density $P({\bf {n}_m},t)$, where $\bf {n}_m$ is the unit vector
describing the direction of the macro-spin moment $\bf m$; in
spherical coordinates ${\bf n_m} $ is characterized by
$(\theta,\phi)$. Before one turns on the magnetic field and the
current, the probability density takes its equilibrium value. For a
uniaxial anisotropy-only situation,
\begin{equation}
P({\bf {n}_m},0) = P_0 \exp (-\xi \sin^2 \theta )
\end{equation}
for $0\leq\theta\leq\pi/2$ and $P({\bf {n}_m},0)=0$ for $\pi/2 <
\theta \leq \pi$, where $P_0$ is the normalization factor
($\int_0^\pi P\sin\theta d\theta=1$), $P_0 \approx 2\xi$, if we
consider $\xi \equiv KV/k_BT \gg 1$ where $K$ is the anisotropy
constant and $V$ is the volume of nanomagnet. To determine $P({\bf
{n}_m},t)$ after one turns on the field and the current at $t=0$, we
need to solve the Fokker-Planck equation given below
\begin{equation}
\frac{\partial P}{\partial t} + \nabla \cdot {\bf J} -D \nabla^2 P =
0
\end{equation}
where ${\bf J}$ is the probability current and $D=\alpha\gamma
k_BT/m$ is the diffusion constant, $\alpha$ is the damping
parameter, $\gamma$ is the gyromagnetic coefficient, and $m=|\bf
m|$. The probability current is
\begin{equation}
{\bf J}= P \frac{d{\bf{n}_m}}{dt} = - \gamma P {\bf{n}_m}\times
[{\bf H}_e + \alpha {\bf {n}_m}\times ({\bf H}_e - {\bf H}_s) ]
\end{equation}
where we have used the Landau-Lifshitz-Gilbert equation including
the spin torque term ${\bf H}_s = (I p )(\hbar/2q)(1/\alpha m){\bf
n}_s$ ($I$ is current density, $p$ is the spin polarization
coefficient, $q$ is the electron charge, and ${\bf n}_s$ is a unit
vector in the direction of the magnetization of the pinned layer).
To make the solution mathematically tractable, we consider the case
where the magnetic field ${\bf H}$ and ${\bf n}_s$ are parallel to
the anisotropy axis, i.e., ${\bf H}_e = (H+H_k \cos\theta) {\bf e}_z
$ and ${\bf n}_s={\bf e}_z$ where $H_k = 2K/M_s$. With these
simplifications, Eq.~(3) reduces to
\begin{equation}
\frac{\partial P}{\partial t} = \frac{\alpha\gamma}{m}\frac{
\partial}{\partial x} \left[ (1-x^2) \left( \frac{\partial
U_e}{\partial x} P +k_BT \frac{\partial P} {\partial x} \right)
\right]
\end{equation}
where $x=\cos\theta$, and we have defined the effective potential
$U_e = (H_s - H)m x -(H_k/2)mx^2$. Equation (4) can be solved via
the method of separation of variables. Let's assume $P(x,t)=
f(t)u(x)$ and it is easy to see that $f(t) = \exp (-\lambda t)$ and
$u(x)$ satisfies
\begin{equation}
\frac{\alpha\gamma}{m}\frac{d}{dx} \left[ (1-x^2) \left( \frac{d
U_e}{d x} u (x) +k_BT \frac{d u (x)}{dx} \right) \right] =-\lambda
u(x)
\end{equation}
The above equation can be converted into a Sturm-Liouville
equation,
\begin{equation}
\frac{d}{dx} \left[ (1-x^2) {\rm e}^{- \beta U_e} \frac{dF}{dx}
\right] + {c } {\rm e}^{-\beta U_e}F =0
\end{equation}
where $F(x) = {\rm e}^{\beta U_e(x)}u(x)$ and $c = \lambda
m/\alpha\gamma k_BT$. Equation (6) is the same equation introduced
by Brown \cite{Brown} except that $U_e$ includes the spin-current
term $H_s$. The original FPE (4) is now reduced to the standard
eigenvalve problem. Namely, we determine eigen-function $F(x)=F_n
(x)$ and eigenvalue $\lambda = \lambda_n$ from Eq.~(6) for
$n=0,1,2,...$, and the general solution of Eq.~(4) is
\begin{equation}
P(x,t)= \sum_{n=0} A_n e^{-\beta U_e(x)} F_n (x) {\rm e}^{-\lambda_n
t}
\end{equation}
where the coefficients in Eq.~(7) are determined by the initial
condition Eq.~(1): $A_n = \int_{-1}^{1} dx P(x,0)F_n(x)$, where we
have chosen the eigenfunction $F_n (x)$ to be weighted normalized:
$\int_{-1}^1 dx {\rm e}^{-\beta U_e (x)} F^2_n (x) = 1$.

The remaining task is to solve eigenvalue $\lambda_n$ (or $c_n$) and
eigenfunction $F_n$. Obviously, $\lambda_0 = 0$ is the lowest
eigenvalue with the corresponding eigenfunction $F_0$ being a
constant. Thus, $u_0 (x) \propto \exp(-\beta U_e)$ describes the
thermal equilibrium state. The smallest nonzero eigenvalue
$\lambda_1$ then determines the slowest decaying or switching speed.

The early theoretical attempts toward solving Eq.~(6) \cite
{Brown,Aharoni,Coffey,Scully, Bertram}had been focused on the cases
where there are two potential wells separated by an energy barrier
and $H<H_k$. These solutions do not apply to the fast switching case
where $H>H_k$. In this case, there is no energy barrier and the only
stable solution is at $x = -1$ or $\theta = \pi$. In the following,
we develop two methods to solve the Eq.~(6): one is the approximate
minimization which provides an upper bound for $\lambda_1$ and the
other is to numerically determine the coefficients of polynomial
series when Eq.~(6) is expanded into polynomials.

The variational method can be obtained by multiplying $F(x)$ to
Eq.(6) and then by integrating from -1 to 1,
\[
\int_{-1}^1 dx F \frac{d}{dx} \left[ (1-x^2) {\rm e}^{- \beta U_e}
\frac{dF}{dx} \right] + c \int_{-1}^1 {\rm e}^{-\beta U_e}F^2 dx =0.
\]
Performing integration by parts of the first term, we find
\begin{equation}
c = {\cal D}/{\cal H}
\end{equation}
where
\begin{equation}
\left\{\begin{array}{ll} {\cal D} = \int_{-1}^1 dx {\rm e}^{-\beta
U_e}(1-x^2) \left( \frac{dF}{dx} \right) ^2 \\\\
{\cal H} = \int_{-1}^1 dx {\rm e}^{-\beta U_e} F^2 (x)
\end{array}\right.
\end{equation}
Since both ${\cal D}$ and ${\cal H}$ are definitely positive, the
variational principle is to choose a trial function $F (x)$ such
that $c$ is minimized, and the non-zero minimum $c$ represents the
upper bound of the lowest eigenvalue $c_1$. Additional constriction
of $F(x)$ is its orthogonality with $F_0$ which is a constant, i.e.,
\begin{equation}
\int_{-1}^1 {\rm e}^{-\beta U_e} F(x) dx = 0.
\end{equation}
Since $\exp(-\beta U_e) $ is sharply peaked at $x = -1$ in our case,
we change the variable as $x = y -1$ and thus $\beta U_e = \eta y-
\xi y^2 + const$, where we define $\eta = \frac{m}{k_BT}(H_s- H +
H_k)$ and $\xi =\frac{m}{k_BT} \frac{1}{2}H_k$. We consider a simple
trial function $F(y)=1+b_1y+b_2y^2$ where $b_1$ and $b_2$ are
variational parameters. By placing $F(y)$ into Eq.~(9) and (10), we
have
\begin{equation}
\left\{\begin{array}{ll} {\cal D}= \int_0^2 y(2-y) {\rm e}^{-\eta y
+ \xi y^2} (b_1+2b_2y)^2 dy \\\\
{\cal H} = \int_0^2 {\rm e}^{-\eta y + \xi y^2} (1+b_1y+b_2y^2)^2 dy
\\\\
\int_0^2 {\rm e}^{-\eta y + \xi y^2} (1+b_1y+b_2y^2)dy =0
\end{array}\right.
\end{equation}
After the completion of these three integrations, the eigenvalue
$c_1$ is obtained by minimizing $\cal D/H$ with respect to $b_2$ or
$b_1$. The above integrations can be analytically obtained if we we
notice that $y$ is limited to $1/\eta$ due to exponential ${\rm
e}^{-\eta y}$ and thus the term $e^{\xi y^2}$ is small and can be
expanded as: ${\rm e}^{\xi y^2} = 1+ \xi y^2 +\frac{1}{2}\xi^2 y^4$.
By using $\int_0^2 y^n {\rm e}^{-\eta y} dy \approx n!/\eta^{n+1} $,
for example, the third equation in Eq.~(11) is
\begin{eqnarray}
\frac{1}{\eta} + \frac{2\xi}{\eta^3} + \frac{12\xi^2}{\eta^5} +
\left( \frac{1}{\eta^2} + \frac{6\xi}{\eta^4} +
\frac{60\xi^2}{\eta^6} \right) b_1 \nonumber\\ + \left(
\frac{2}{\eta^3} + \frac{24\xi}{\eta^5} +
\frac{360\xi^2}{\eta^7}\right) b_2 =0
\end{eqnarray}
Note that the leading orders of $b_1$ and $b_2$ are $\eta$ and
$\eta^2$ respectively, and the above equation is valid up to
$\eta^{-3}$. If we consider the fast switching case, i.e.
$\eta\gg\xi$, or $H_s -H \gg H_k$, the equation (12) can be further
approximated as
\begin{equation}
b_1 =  -2b_2\eta^{-1}- \eta
\end{equation}
We can similarly calculate $\cal H$ and $\cal D$. Substituting $b_1$
by Eq.~(13), the expression of $\cal D/H$ contains only $b_2$. Then,
we find $b_2$ by minimizing $\cal D/H$
\begin{equation}
\frac{d}{db_2} \left( \frac{\cal D}{\cal H} \right) = 0
\end{equation}
The results are:
\begin{equation}
\left\{\begin{array}{lll}
b_2 = \frac{1}{2}\eta^2, &  b_1 = -2\eta &\nonumber\\\\
{\cal H} = \frac{1}{\eta}, & {\cal D} = 4, &  c_1 = \frac{\cal
D}{\cal H} = 4\eta\nonumber
\end{array}\right.
\end{equation}
Rewriting in terms of original parameters, we find the slowest
decaying rate is
\begin{equation}
\lambda_1 = 4 \alpha\gamma  \left( H_s -H\right) = 4 (I/I_c
-1)\tau_0^{-1}
\end{equation}
where $\tau_0^{-1} = \frac{p\mu_B}{qm}I_c = \alpha\gamma(H+H_k)$.

Although the variational method is able to analytically obtain the
upper bound for the rate of the relaxation, its accuracy is unknown.
A more rigorous method is to expand $F(x)$ via Taylor series, i.e.,
$F(x) = \sum a_nx^n$ ($n = 0,1,2...$). By substituting it into
Eq.~(6),we find the recurrence formula for coefficient $a_n$
\begin{eqnarray}
(n+2)(n+1)a_{n+2}-\zeta(n+1)a_{n+1} \nonumber\\
+[c_n-n(n-1)+n(2\xi-2)]a_n \nonumber\\ + \zeta(n-1)a_{n-1} -
2\xi(n-2)a_{n-2}=0
\end{eqnarray}
where $\zeta = \eta -2\xi = \frac{m}{k_BT}(H_s- H)$. In the case of
$\zeta=\xi=0$, the above equation reduces to a two-term recursion
formula, and the solutions are the Legendre polynomials with $c_n =
n(n+1)$. This is the well-known solution for the thermal decay of a
free particle \cite{Brown}. In the case of $\zeta=0$, Eq.~(16)
reduces to a three-term recursion formula, and the continued
fraction method can be used to obtain the smallest nonzero
eigenvalue $c_1 = 2-\frac{4}{5}\xi + O(\xi^2)$\cite{Aharoni}.

In our case where both $\zeta$ and $\xi$ are non-zero, above methods
can not be used directly. One way to solve the problem is to
numerically calculate Eq.~(16) by simply keeping enough terms (i.e.
up to a large number $N = 300$) to ensure the convergence of $c_1$.
In Fig.~1,we show that the relaxation rates as a function of
temperature ($H_k$ is fixed) for different $I/I_c (\geq 1)$. In
Fig.~2, we show that the relaxation rate is linearly dependent of
$I/I_c - 1$ when $I/I_c > 1.5$. By comparing with Eq.~(15) derived
from the variational approach, we find that to a good approximation
the relaxation rate is a factor of two smaller, i.e.,
\begin{equation} \lambda_1 = 2 (I/I_c -1)\tau_0^{-1}. \end{equation}

\section{Comparison with Sun's model}

Sun's model \cite{Sun1} assumes that thermally distributed initial
magnetization states determine the distribution of switching time in
case of $I>I_c$. In this model, the switching time at zero
temperature is estimated as $\tau = \tau_0(I/I_c - 1)^{-1}
\ln(\pi/2\theta)$, where $\theta$ is the initial angle of
magnetization whose distribution is given by Eq.~(1). The
corresponding distribution of switching time is calculated from the
definition: $D(\tau)d\tau = - P(\theta)\sin\theta d\theta$. By
defining the switching probability density $D(\tau)d\tau = - P({\bf
n}_{\bf m} , 0)\sin\theta d\theta$ and utilizing the above relation
between $\tau$ and $\theta$, the probability of not being switched
is
\begin{equation} E_r(t) \equiv 1
- \int_0^tD(\tau)d\tau \simeq
\frac{\pi^2\xi}{4}\exp\left(-\frac{2(I/I_c - 1)}{\tau_0}t \right)
\end{equation}
where a long time limit is taken. The relaxation rate
is identical with Eq.~(17). This suggests the initial-condition
randomization is the leading cause for switching time distribution in
the limit of large $\xi$ and $I/I_c \gg 1$. Deviation
occurs when $\xi$ is small, such as shown in
Fig.~1, or when the current is near $I_c$, as in Fig.~2.

This work is supported in part by IBM.

\pagebreak
\bigskip
\noindent {Figure Caption}

\bigskip
\noindent{FIG.1}(Color online) Numerical results of the smallest
nonzero eigenvalues of Eq.~(16) for increasing $\xi$, in cases of
different $I/I_c$. $\tau_0^{-1} = \alpha\gamma(H+H_k)$.
\bigskip

\bigskip
\noindent{FIG.2}(Color online) Numerical results of the smallest
nonzero eigenvalues of Eq.~(16) for increasing $I/I_c - 1$, where
$\xi = 80$. The red line is the fitting line: $\lambda_1\tau_0 =
2(I/I_c - 1)$.
\bigskip
\newpage

\begin{figure}
\centering
\includegraphics[width=12cm]{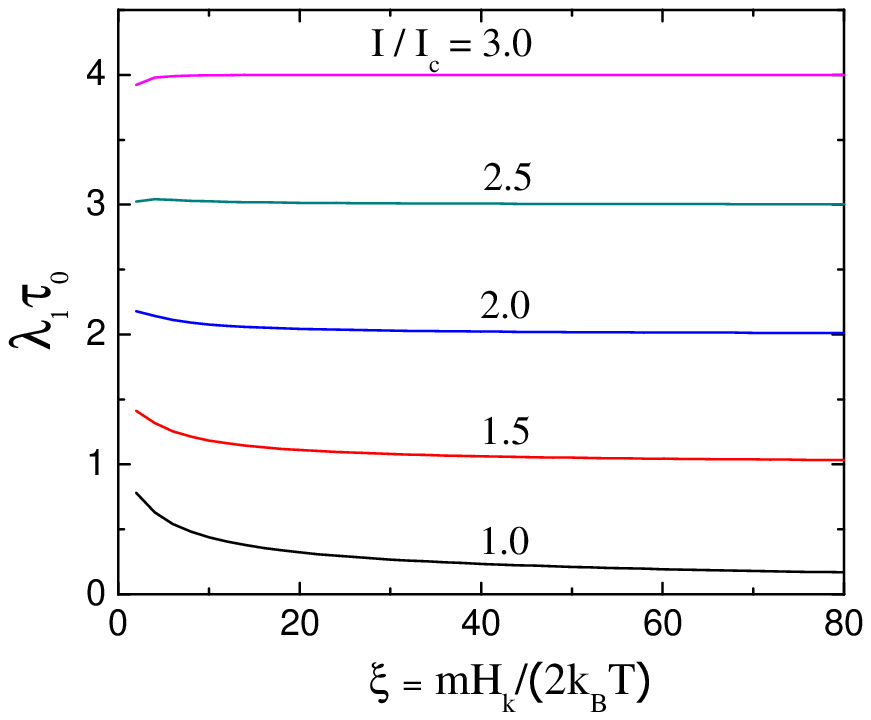}
\caption{}
\end{figure}
\bigskip
\pagebreak
\newpage
\begin{figure}
\centering
\includegraphics[width=12cm]{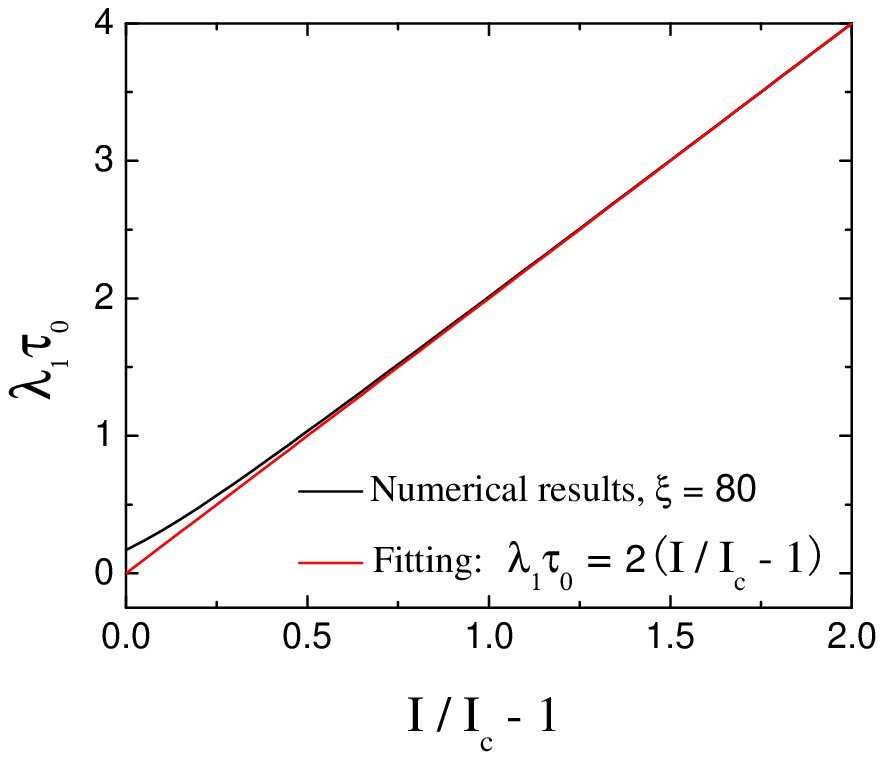}
\caption{}
\end{figure}

\end{document}